\documentclass[nofootinbib,12pt,pra,notitlepage,fleqn]{revtex4}
\usepackage{bm}
\usepackage{amssymb}
\usepackage{graphicx,xcolor}
\usepackage{amsmath,graphicx}
\usepackage{moresize}
\usepackage{epigraph}
\renewcommand{\epigraphsize}{\tiny}
\usepackage{csquotes}
\usepackage{amsmath}
\usepackage{amsfonts}
\usepackage{graphicx}
\def\la{{\langle}}
\def\u{\hat U}

\def\x{\overline}

\def\D{\overline{D}}
\def\ot{\otimes}

\def\W{\overline W}

\def\e{\enquote}

\def\q{\quad}

\def\n{\\ \nonumber}
\def\ra{{\rangle}}

\def\Pr{{Pr'}}
\def\Pr{{Pr'}}
\newcommand{\br}[1]{\textcolor{black}{#1}}

\begin{document}

\title{Wigner's friend scenarios and the internal consistency of
the standard quantum mechanics }
\author{D. Sokolovski$^{a,b}$}
\email{dgsokol15@gmail.com}
\author{A. Matzkin$^{c}$}
\affiliation{$^{a}$ Departmento de Qu\'imica-F\'isica, Universidad del Pa\' is Vasco,
UPV/EHU, Leioa, Spain}
\affiliation{$^{b}$ IKERBASQUE, Basque Foundation for Science, E-48011 Bilbao, Spain}
\affiliation{$^{c}$ Laboratoire de Physique Th\'eorique et Mod\'elisation, CNRS Unit\'e
8089, CY Cergy Paris Universit\'e, 95302 Cergy-Pontoise cedex, France}
 \date\today
\begin{abstract}
The extended Wigner's friend problem deals with two Observers each measuring
a sealed laboratory in which a friend is making a quantum measurement. We
investigate this problem by relying on the basic rules of quantum mechanics
as exposed by Feynman in the well-known ``Feynman Lectures on Physics''.
Although recent discussions have suggested that the extended Wigner's friend
problem cannot consistently be described by quantum theory, we show here
that a straightforward application of these standard 
 rules
results in a non-ambiguous and consistent account of the measurement
outcomes for all agents involved.
\end{abstract}
\maketitle
\setlength\oddsidemargin{-1cm}
\setlength\textheight{22cm}
\setlength\topmargin{-1cm}
\setlength\textwidth{18cm}
\renewcommand{\epigraphsize}{\tiny} \setlength{\epigraphwidth}{0.5\textwidth}

\let\originalepigraph\epigraph
\renewcommand\epigraph[2]{\originalepigraph{\textit{#1}}{\textsc{#2}}} 

\section{Introduction}

The celebrated Wigner Friend thought experiment \cite{wigner} sought to expose
the possible ambiguities of quantum theory when it is applied to
measurements made by intelligent agents. In this thought experiment, an Observer is requested to
measure a sealed laboratory in which a Friend is making a measurement. As
Wigner argued, if the laboratory contained only an atom going through the
inhomogeneous magnetic field of a Stern-Gerlach device, the laboratory's
evolution would be unitary. But if a Friend is inside and actually measures
the spin, should the Observer, who assumes quantum theory applies to systems
of any scale, still describe the situation by unitary evolution ?

Recently, an extended version of this Wigner Friend setup was
proposed \cite{renner,bruckner}. This extended version involves two external
Observers each measuring a different laboratory, containing a Friend who is measuring
a spin. 
\ According
to Frauchiger and Renner \cite{renner}, whose setup employed non-maximally
entangled spin states, quantum mechanics becomes inconsistent in the following sense.
An inference based on the Friends' observed outcomes implies that the external Observers
should never obtain a particular set of outcomes since the corresponding
probability will vanish. 
However, a direct computation which ignores the
Friends' outcomes indicates that the same probability \emph{is not} zero.
In \cite{bruckner}  Brukner, who employed maximally entangled spin states, concluded that the
facts (as defined by the agents' measurements) cannot be the same for each
Observer and his associated Friend.\ This conclusion is reached by setting
up an inequality involving correlations between the Friends' and the external
Observers' outcomes, such that it may be obeyed only if these outcomes are
observer-independent. It is then shown that the inequality is violated, 
if quantum mechanics is applied.

Several authors have scrutinised both the results reported in Refs. \cite%
{renner,bruckner} and the implicit or explicit assumptions made in deriving
them. Some of these works analysed the extended Wigner's friend setup within frameworks arising in
specific interpretations of quantum mechanics, in which the prescriptions for
computing the probabilities of the measured outcomes are different from those
 followed by Frauchiger and Renner, or by Brukner.
Such studies employed different variants of the de Broglie-Bohm interpretation 
\cite{sudbery,lazarovici},  the consistent histories formalism \cite%
{lombardi}, or the decoherence approach \cite{relano,2WF8}.
The Bayesian and relational  approaches to quantum mechanics  were used 
for the same purpose in  \cite{fuchs} and  \cite{relational}. 
Yet more elaborate methods included deriving
 theory-independent inequalities 
assuming universal validity of all events to all agents \cite%
{wiseman}, or adding non-invasive weak measurements in order to allow the external
Observers monitor the state of the Friends' laboratories \cite{WFS-WM}.
Translating the Wigner friend setup into an interferometric setup where the external
Observers actions may erase a photon's memory state was considered in \cite{andrew}.
A classical analogy of the scenario was proposed and studied in \cite{Lost}.
\newline
Among the viewpoints thus expressed,
still missing is a standard textbook account of what happens
in extended Wigner Friend scenarios. 
Arguably, the approach pioneered by Feynman in his texts \cite{FeynL,FH} is  the 
most basic pragmatic implementation of the standard  quantum theory. \br{Feynman 
	gives a set of rules stating how probabilities should be computed from the amplitudes, viz. through a ``sum over virtual paths'' (involving amplitudes) or a ``sum over real paths'' (involving probabilities).}
Recently \cite{WFS-EPL}, we applied the Feynman's rules for adding the 
probability amplitudes to the original Wigner Friend experiment. 
In \cite{WFS-EPL} a coherent narrative was constructed by taking into account 
stable material records produced by each act of observation. 
In this paper we extend the approach to more complex scenarios, in order to 
look for inconsistencies, similar to those discussed in \cite{renner}-\cite{andrew}.
We will ask whether similar problems 
would also arise in the frame work proposed in \cite{FeynL,FH}, and if they would, then in 
what particular form? \br{We will see that the rules given by Feynman limit the {\it apriori} assumptions which can be made about Wigner-Friend scenarios, since 
some assumptions will violate these rules. This is because the existence of the probabilities of measurement outcomes relies on the existence of the corresponding material records. }	
\newline
The rest of the paper is organized as follows. 
In Sec. \ref{res} we  summarize
rules given by Feynman. In Sec. II we apply the rules to the 
original Wigner friend problem.
Section III discusses a more complex scenario, involving a total 
of four Observers. In Sect. V we show how one can 
(although by no means needs to) arrive at a logical contradiction. 
In Sect.VI we resolve the contradiction by
straightforward application of Feynman's prescriptions  \cite{FeynL,FH}. 
Our conclusions are in Sec. \ref{conclu}.
Additional relevant details will be found in several Appendices.


\section{Feynman rules \label{res}}

For the purpose of our discussion, the general rules of \cite{FeynL} (see
also Ch.\ 1 of \cite{FH}) can be condensed into the following propositions.%
\newline
A) Quantum mechanics provides probabilities of the outcomes of a series of
measurements accessible, at least in principle, to an Observer or Observers. 
\newline
B) Probability is computed as the absolute square of a complex valued
probability amplitude, {\it at the end} of the experiment. \newline
C) If a route consists of several parts, the amplitudes of the parts must be
multiplied. \newline
D) The amplitudes are added if several routes leading to the same outcome
are not distinguished by an experiment, even {\it in principle.} \newline
E) The amplitudes are never added for different and distinct final states.%
\newline
F) The amplitudes are never added if information about the final state is in
principle available, regardless of whether this information is known or
used. 
\newline
\newline
These principles are readily applied to an experiment consisting of several
measurements, made one after another on the same quantum system. Note that
when considering a series of outcomes, the resulting probabilities can be
obtained with the help of amplitudes evaluated for all relevant virtual
(Feynman) paths connecting the states in the Hilbert space of the measured
system.
 Note also that different sets of measurements may lead to different
(incompatible) statistical ensembles. 
{\color{black}
\section{One Wigner and one Friend} \label{original}
In the original Wigner Friend scenario (WFS) Wigner \cite{wigner} considers
an Observer, say $W$, observing a friend, $F$.
Using  a device $D$, $F$
measures  a two-level  system, initially in a state $|s_0\ra$,  in a basis $|up\ra$ and $|down\ra$
The possible states
of the system (S), device (D), and the memory ($\mu$), therefore,  are
\begin{equation}\label {s1}
|i\ra \ot |D(i)\ra\ot |\mu(i)\ra\, \q i=up, down
\end{equation}

In \cite{WFS} Wigner argued that a linear superposition
\begin{equation} \label{zw1}
\sum_{i}\left\langle i\right\vert \left. s_{0}\right\rangle \left\vert i\right\rangle \ot \left\vert D(i)\right\rangle\ot  \left\vert \mu (i)\right\rangle 
\end{equation}
could not represent F's situation once his 
measurement is completed, 
since Eq.(\ref{zw1}) appears to imply an uncertainty regarding $F$'s outcome, resolved only
 after W measures the composite \{S+D+$\mu$\}. 
 The question of whether Eq. (\ref{zw1}) should  be correct from W's viewpoint seemed to remain open, assuming W is able to coherently control the composite. 
\newline
A different account of Wigner's experiment can be given with the help of Feynman rules of Sect.II. 
As in the example in Appendix B, it is sufficient to  consider virtual scenarios for the system, with an understanding (rule D) that their amplitudes, 
 or probabilities must be added,
 depending on whether the 
scenarios can or cannot be distinguished. 
To distinguish between possible scenarios one would require a record of the system's earlier condition
( a memory, a note, or an identical copy of the system) to be available when the experiment is finished, and the results are obtained (cf. rule B).
By the rule D, it does not matter whether all the records were inspected by the Observer(s), since it is their mere existence which determines
how the amplitudes should be handled. 
It is important to note that if $W$'s action succeeds in erasing all records made by $F$ earlier, then, by the rule D, 
all information about the state of the system at the time of $F$'s measurement should be irretrievably lost to interference.

Elaborating on Wigner's original scenario, let us consider the following situations, similar to those discussed in detail in Appendix B.

I) W measures only the system in a different orthogonal basis, $|fail\ra=\alpha |up\ra + \beta |down\ra$ and $|ok\ra=\gamma |up\ra + \delta|down\ra$, so  $F$'s  memory and device remain untouched. 
At the end of the experiment, i.e., just after $W$'s measurement, there are several possibilities of recovering $F$'s outcome.
For example, W  could i) look at $F$'s probe,  ii) ask $F$, who will the have to consult his memory, or 
iii) ask $F$ to look at his device again.  The probabilities of the four possible outcomes are, therefore, given by 
\begin{equation}\label {s2}
P'(j,i)=|A^S(j\gets i\gets s_0) |^2, i=up, down, \q j= fail, ok.
\end{equation}
and 
\begin{equation}\label {s3}
P'(j)=\sum_i P(j,i). 
\end{equation}
\br{$A^S(j\gets s_0)=\la j| U(t,t_0)|s_0 \ra$ is generically the  transition amplitude connecting an initial state $|s_0\ra$ at time $t_0$ to a state $|j \ra$ at time $t$, where $U$ is the system evolution operator. Following rule C), the amplitude if an intermediate state $|i \ra$ is chosen is given by 
	$A^S(j\gets i\gets s_0)=\la j|i\ra\la i |s_0\ra$ where we assume the system has no own dynamics.}
Note that by F) of the previous Section, Eq.(\ref{s3}) holds even if the  inquiry about $F$'s outcome was never made, it suffices that 
in could be made {\it in principle}. Note also that $W$ can also measure the joint systems  $\{S+D\}$, $\{S+\mu\}$, and $\{D+\mu\}$
in any orthogonal basis, e.g., $|Fail\ra=\alpha |D(up)\ra\ot |up\ra + \beta|D(down)\ra\ot |down\ra$ and $|Ok\ra=\gamma |D(up)\ra\ot |up\ra + \delta|D(down)\ra\ot |down\ra$, etc. Equations (\ref{s2})-(\ref{s3}) would be continue to be valid for as long as at least one record, 
carried by $\mu$, $D$, or $S$, is available once the experiment is finished. This is essentially the case b) of the Appendix B.


II) Assume now $W$ decides to measure the composite  $\{S+D+\mu\}$ in a basis 
\begin{eqnarray}\label {s4}
|Fail\ra=\alpha |\mu(up)\ra\ot|D(up)\ra\ot |up\ra + \beta|\mu(down)\ra\ot|D(down)\ra\ot |down\ra, \n
|Ok\ra=\gamma |\mu(up)\ra\ot |D(up)\ra\ot |up\ra + \delta|\mu(down)\ra\ot|D(down)\ra\ot |down\ra\q.
\end{eqnarray}
In this situations, all records of $F$'s outcome would be erased, and the system's virtual paths $\{j \gets up \gets s_0\}$ and $\{j \gets down \gets s_0\}$, 
$j=fail, ok$ cannot be distinguished. The only  outcomes are those obtained by $W$ and, in accordance with Rule D) of the previous Section,  
we have 
\begin{equation}\label {s5}
P''(j)=|A^S(j\gets up \gets s_0)+A^S(j\gets down \gets s_0) |^2, j=Fail, Ok.
\end{equation}
This is, essentially the case c) of the Appendix B. 

In summary, if we define a Wigner's Friend setup by the requirement that both F and W make full measurements with outcomes confirmed by material records at the end, then it is represented by the scenario I).
The case II) is, by necessity,  different since the records, duly produced by $F$ earlier, do not survive until the end of the experiment. 
In both cases application of the Feynman rules given above leads to an unambiguous answer.
The probabilities, $P'(j)$ and $P''(j)$ of cases I) and II) are different, because they refer to different measurements made by $W$.
Disagreements of this type will become more pronounced in a more complex extended Wigner friend scenario, which we discuss next. }

\section{Two Wigners and two Friends}
In a more complex Wigner's friend scenario (2W2FS), considered
in
\cite{renner}, there are two 2-level systems, called a "coin", and a "spin",
and four participants: a pair of friends, $\overline{F}$ and $F$, and
two external Observers, $\overline{W}$ and $W$. $\overline{F}$  measures the coin (C), and $F$ measures the spin (S), each in
her/his own basis. 
The orthonormal measurement basis of the
coin consists of states  $|heads{\rangle }$ and $|tails{\rangle }$. The spin states
in $F$'s measurement basis will be labeled $|up{\rangle }$ and $|down{%
\rangle }$. The Friends \e{laboratories} ($L$, $\x {L}$) consist of all material devices and objects, involved in the measurements ($D$, $\x D$),  
 (initially in states $|D(0)\ra$ and $|\x D(0)\ra$) plus the coin and the spin, respectively.
 In order to simplify the already cumbersome notations, we included $F$ and $\x F$'s memories 
 into what we now call $D$ and $\x D$ \br{(note that strictly speaking, we only need the Friends' memories or records to be inside their respective laboratories as this is what quantum mechanics applies to; this can be rendered metaphorically as $ F$ and $\x F$ and $F$ being inside $L$ and $\x {L}$).}
A measurement entangles the contents of a lab with the coin or the spin according to
 \begin{eqnarray} \label{0001}
 |\x D(0)\ra\ot |\phi^C\ra \to \la heads|\phi^C\ra |\x D(heads)\ra\ot |heads\ra+\la tails|\phi^C\ra |\x D(tails)\ra\ot |tails\ra\n
\equiv  \la heads|\phi^C\ra |\x L(heads)\ra+\la tails|\phi^C\ra |\x L(tails)\ra
,\n
  | D(0)\ra\ot |\phi^S\ra \to \la up|\phi^S\ra | D(up)\ra\ot |up\ra+\la down|\phi^S\ra |\x D(down)\ra\ot |down\ra\n
  \equiv  \la heads|\phi^S\ra |L(up)\ra+\la down|\phi^C\ra | L(down)\ra,
 \end{eqnarray}
where $|\phi^C\ra$ and  $|\phi^S\ra$ are arbitrary states of the coin and the spin, 
and $|\x L(has/tails)\ra$ and $ | L(up/down)\ra$ are the states of the entire laboratories. 
\newline
The development is as follows.
Neither coin, nor spin, have their own dynamics. At $t=t_0$ the coin, the spin, and the two labs are prepared in a state
\begin{equation}\label{0002}
|\Psi _{0}{\rangle }=[|heads{\rangle }+\sqrt{2}|tails{\rangle }]/\sqrt{3}%
\otimes |down{\rangle }\otimes |\x{D}(0)\ra \ot |D(0)\ra.
\end{equation}%
Then, at $t_1>t_0$, $\x F$'s device is coupled to the coin according to (\ref{0001}).
At a $\tau > t_1$ the spin and coin coupled by means of a
Hamiltonian $\hat{H}_{int}=i(\pi /4)|tails{\rangle }{\langle }tails|\otimes
\lbrack |up{\rangle }{\langle }down|-|down{\rangle }{\langle }up|]\delta
(t-\tau+\epsilon/2 )$, so that the corresponding evolution operator  is 
\begin{eqnarray}\label{0003}
\hat{U}^{c+s}(\tau,\tau-\epsilon)[|heads{\rangle }+\sqrt{2}|tails{\rangle }]/\sqrt{3}
\otimes |down{\rangle }\q\q\q\q\q\q\q\q\q\q\q\q\q\q\q\n
=|heads{\rangle }{\langle }heads|\otimes \hat{1}%
_{spin}+|tails{\rangle }{\langle }tails|\otimes \lbrack \hat{1}_{spin}+|up{%
\rangle }{\langle }down|-|down{\rangle }{\langle }up|]/\sqrt{2}.\quad 
\end{eqnarray}%
One can say that
$\overline{F}$ sends $F$ the spin in a state $|down\rangle $ if his/her outcome
is $Heads$, or in a superposition $(|up\rangle +|down\rangle )/\sqrt{2}$, if
it is $Tails$.
At $t_2 > \tau$, $F$'s device is coupled to the spin according to (\ref{0001}).
\newline
Finally, at a $t_3 > t_2$, $\x W$ and $W$ measure entire $\x F$'s and $F$'s laboratories in the bases
 \begin{eqnarray} \label{0004}
|\x {fail}\ra=[|\x L(heads)\ra+|\x L(tails)\ra]/\sqrt 2, \q |\x {ok}\ra=[|\x L(heads)\ra-|\x L(tails)\ra]/\sqrt 2,
 \end{eqnarray}
and
 \begin{eqnarray} \label{0005}
|{fail}\ra=[| L(up)\ra+|L(down)\ra]/\sqrt 2, \q |{ok}\ra=[| L(up)\ra-|L(down)\ra]/\sqrt 2,
 \end{eqnarray}
using 
couplings similar to those in Eq.(\ref{0001}).

\section{A contradiction}
The apparent contradiction associated with the scenario of the previous Section, put forward in \cite{renner}, can be summed up in the following manner \cite{WFS-WM}.
By $t=t_3$, a unitary evolution including 
the couplings (\ref{0001}) and (\ref{0003}) converts the initial state (\ref{0002}) into
\begin{equation}  \label{C1}
|\Phi(t_3)\ra=[|\x L(heads)\ra\ot|L(down){\ra}+|\x L(tails){\ra}\ot|L(up){\ra}+|\x L (tails){\ra }\ot|L(down){\ra}]/\sqrt{3}.
\end{equation}
It does not contain a term $|\x L (heads){\ra }\ot|L(down){\ra}$, so the likelihood of $\x F$ and $F$ jointly obtaining the outcomes
$Heads$ and $Up$ is zero, 
\begin{eqnarray} \label{C2}
P(Heads,Up) =0, 
\end{eqnarray}
while, obviously,
\begin{eqnarray} \label{C2a}
P(Heads,Down) =P(Tails,Down)= P(Tails,Up)=1/3.
\end{eqnarray}
\newline 
However, in another equivalent representation the state (\ref{C1}) misses a term $|ok\ra\ot |tails\ra$
\begin{equation}  \label{C3}
|\Phi(t_3)\ra=[|\x L(heads)\ra\ot|fail{\ra}-|\x L(heads){\ra}\ot|ok{\ra}+2 |\x L (tails){\ra }\ot|fail{\ra}]/\sqrt{6}, 
\end{equation}
so that 
\begin{eqnarray} \label{C4}
P(Tails,Ok) =0.
\end{eqnarray}
In yet another representation one has  
\begin{equation}  \label{C5}
|\Phi(t_3)\ra=[2|down{\ra}\ot|\x {fail}\ra+|L(up){\ra}\ot|\x {fail}\ra+2 |L (up){\ra }\ot|\x {ok}{\ra}]/\sqrt{6}, 
\end{equation}
and 
\begin{eqnarray} \label{C6}
P(Down,\x{Ok}) =0.
\end{eqnarray}
Finally, direct calculation of a  probability $P(Ok,\x{Ok})$ yields
\begin{eqnarray} \label{C7}
P(Ok,\x{Ok}) =|{\langle}ok|{\langle}\x{ok}|\Phi(t_3){\rangle }|^2=1/12.
\end{eqnarray}
Now the following reasoning leads to a contradiction, 

(i) By Eqs.(\ref{C6}), ($\overline{W}$'s) result "$\overline{Ok}$" implies an "$%
Up$" result for $F$'s spin.

(ii)  By Eq.(\ref{C4}) ($W$'s) result "${Ok}$" implies a "$Heads$"
result for $\bar{F}$'s. 

(iii) But, by Eqs.(\ref{C2}) an "$Up$" result for the spin implies a "$Tails$" result for the coin,
but that is not true, according to Eq.(\ref{C7}).
 \newline
Hence,

(iv) $\bar{W}$ and $W$ will never obtain their "$\overline{Ok}$" and "$Ok$"
results at the same time. 
\newline
However, Eq.(\ref{C7}) shows that  conclusion (iv) is wrong.
\newline
One way to deal with the contradiction is to treat all apparently legitimate  results (\ref{C2}), (\ref{C4}), and (\ref{C6}) 
as the participants' personal experiences (observer-dependent facts \cite{bruckner})  which should not be compared against each other.
\newline
Another possibility is to select the right answer for the 2W2F scenario of Sect. IV
in terms of the available records, and find out which scenarios can be attached 
to the statements (i)-(iii). 

Next we will do so with the help of the Feynman's rules of Sect.II

\section{An analysis}
{\color{black}
Our analysis of the 2W2FS of Sect.IV  will be similar to that of the WFS Sect.III.
 As discussed in Sec. \ref{original}, the Feynman rules rely on the existence material records for measurement outcomes. 
 If all records are retained, rules E) and F) imply a sum of probabilities over the \e{real} paths followed by the system.
 In the absence of such records, rule D) of Sec. \ref{res} prescribes a sum of probability amplitudes over the system's virtual paths. 
In order to clarify the origin of the conflicting statements (\ref{COND3}), (\ref{COND2}),  and (\ref{COND1})
at the heart of the contradiction in Sect.V 
we will consider several cases where $\x W$ or $W$ either leave 
 $\x F$'s or $F$'s records untouched, or choose to erase these records.
 In particular, if $\x W$ chooses to measure only the coin, 
 we can still use Eq.(\ref{0005}), provided we redefine the states  $|fail\ra$ and $|ok\ra$ by replacing
$L(up/down)\ra=|D(up/down)\ra\ot|up/down\ra$ with 
\begin{eqnarray}\label{pros1}
|L(up){\rangle }\equiv |up\ra,\q |L(down){\rangle }\equiv  |down\ra, \q
\end{eqnarray}
it will be possible to deduce the coin's condition at $t=t_1$ from the condition of $\x D$  at $t=t_3$, and we will say that {\it $F$'s record is preserved}
 \newline
 Similarly, $\x{W}$ may choose to {\it preserve $\x{F}$'s record}, by redefining the states $|\x L(heads/tails)\ra$ to be
\begin{eqnarray}\label{pros2}
|\x L(heads){\rangle }\equiv |heads\ra,\q
 |\x L(tails){\rangle }\equiv |tails\ra.\q\q\q
\end{eqnarray}
 As in Sect. III, its is sufficient to consider the virtual scenarios in
the four dimensional Hilbert space of the joint system $\{coin+spin\}$, 
 a welcome specification, given the number of the variables involved.
 }
 \subsection{The system's virtual paths virtual paths}
\begin{figure}[tb]
\includegraphics[angle=0,width=12cm, height= 9cm]{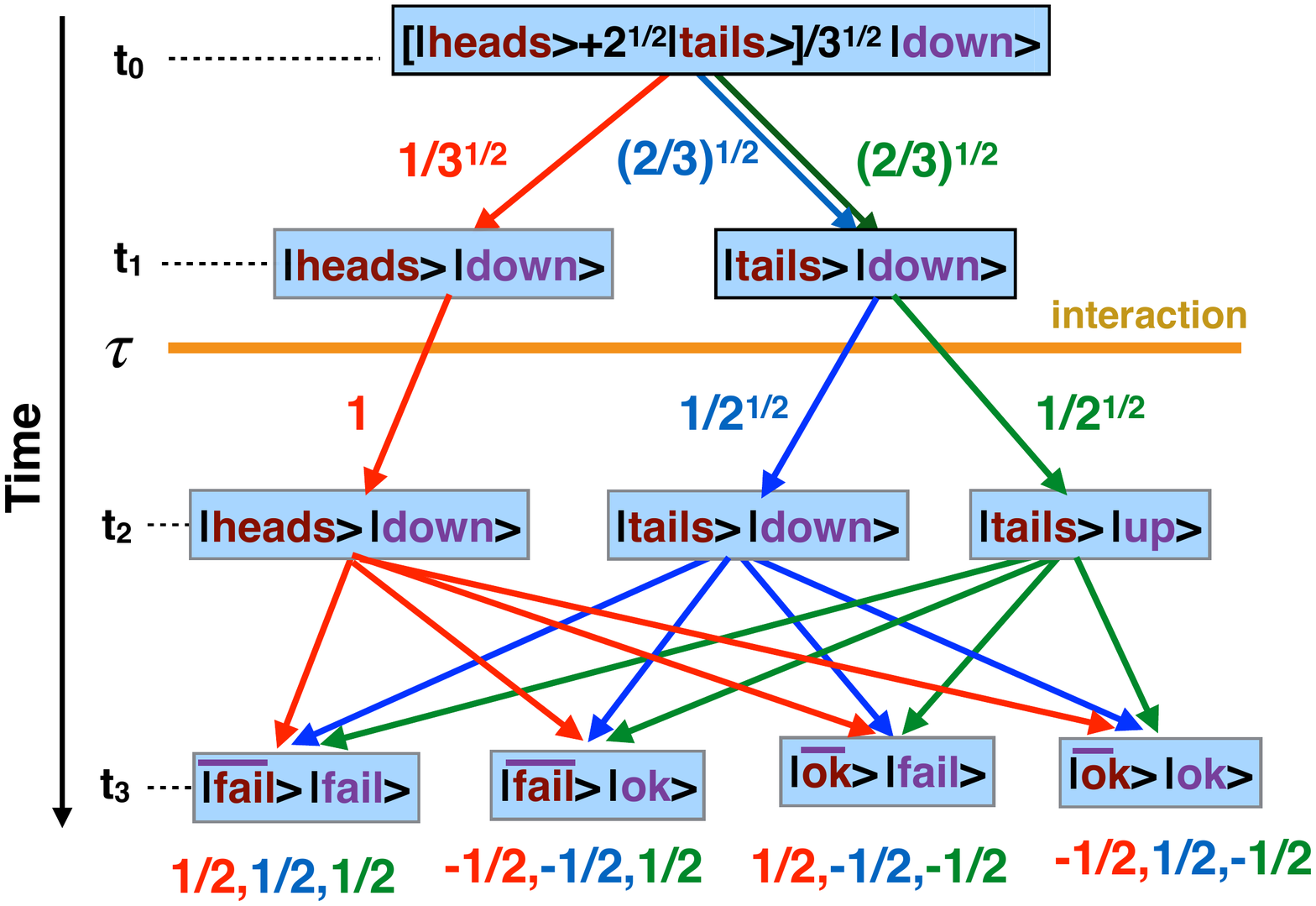}
\caption{Virtual paths in the Hilbert space of the
composite $\{coin+spin\}$ (only the ones with non-zero amplitudes are
shown). $\overline{F}$ measures the coin 
$t=t_{1}$, $F$ does the same at $t=t_{2}$,
and both $\overline{W}$ and $W$ measure at $t=t_{3}$. 
The coupling (\ref{0003}) between the coin and the spin
occurs at $t=\protect\tau $.
The number next to segment of a path, or below it, is the value of
the matrix element of the evolution operator. The full path amplitude is a
product of all such numbers. }
\label{fig:FIG1}
\end{figure}

There are twelve virtual paths of the composite $\{coin+spin\}$ with non-zero amplitudes.
The probability amplitudes for each of the paths shown in Fig.1,  are given by
\begin{align} \label{D2}
A_1\equiv A(\x{fail}\ot fail \gets heads\ot down\gets heads\ot down  \gets \Phi_0)=1/\sqrt{12}, \n
A_2\equiv A(\x{fail}\ot fail \gets tails\ot down\gets tails\ot down  \gets \Phi_0)=1/\sqrt{12},
 \n
A_3\equiv A(\x{fail}\ot fail \gets tails\ot up \gets\ tails\ot down  \gets \Phi_0)=1/\sqrt{12},  
\n
A_4\equiv A(\x{fail}\ot ok\gets heads\ot down \gets heads\ot down \gets \Phi_0)=-1/\sqrt{12}, 
 \n
A_5\equiv A(\x{fail}\ot ok\gets tails\ot down\gets tails\ot down \gets \Phi_0)=-1/\sqrt{12},
\n
A_6\equiv A(\x{fail}\ot ok \gets tails\ot up\gets tails,down\ot   \gets \Phi_0)=1/\sqrt{12}, 
\n
A_7\equiv A(\x{ok}\ot fail\gets heads\ot  down\gets heads\ot down \gets \Phi_0)=1/\sqrt{12}, 
 \n
A_8\equiv A(\x{ok}\ot fail\gets tails\ot down \gets tails\ot down  \gets \Phi_0)=-1/\sqrt{12}, 
 \n
A_9\equiv A(\x{ok}\ot fail \gets tails\ot up \gets tails\ot down  \gets \Phi_0)=-1/\sqrt{12},  
\n
A_{10}\equiv A(\x{ok}\ot ok \gets heads\ot  down \gets heads\ot down  \gets \Phi_0)=-1/\sqrt{12}, 
\n
A_{11}\equiv A(\x{ok}\ot ok \gets tails\ot  down \gets tails\ot down  \gets \Phi_0)=1/\sqrt{12}, 
\n
A_{12}\equiv A(\x{ok}\ot ok \gets tails\ot  up \gets tails\ot down  \gets \Phi_0)=-1/\sqrt{12}, 
\end{align}
where $|\Phi _{0}{\rangle }=[|heads{\rangle }+\sqrt{2}|tails{\rangle }]/\sqrt{3}%
\otimes |down{\rangle }$.
Calculation of the probabilities for all choices made by $\x W$ and $W$ is now reduced to a simple book keeping exercise. 
\subsection{Both $\bar{F}$'s and $F$'s records are erased by $\x W$ and $W$}
We begin with  the scenario where  $\bar{W}$ nor $W$ choose
the bases $\x L(heads/tails)\ra=|\x D(heads/tails)\ra\ot|heads/tails\ra$ and $L(up/down)\ra=|D(up/down)\ra\ot|up/down\ra$ for their respective 
measurements. With the records of $\x F$'s and $F$'s record destroyed, the situation is similar to the one discussed 
in Sect.II.A [cf. Eqs. (\ref{w5})-(\ref{w8})]. In Fig. 1, all
paths leading to the same final state interfere, and following the rules of
 Sec. \ref{res} and Appendix A, we sum the corresponding amplitudes to obtain, 
\begin{eqnarray}  \label{alldest} 
P(\overline{Fail},{Fail})=|A_{1}+A_{2}+A_{3}|^{2}=9/12, \q \n
P(\overline{Fail},{Ok})=|A_{4}+A_{5}+A_{6}|^{2}=1/12,\quad  \notag \\
P(\overline{Ok},{Fail})=|A_{7}+A_{8}+A_{9}|^{2}=1/12,\quad  \notag \\
P(\overline{Ok},{Ok}) =|A_{10}+A_{11}+A_{12}|^{2}=1/12.\quad  \notag
\end{eqnarray}
The corresponding statistical ensemble, shown in Fig. 2a, consists of four real
paths connecting a successful preparation with  $\bar{W}$'s and $W$'s
outcomes and we recover Eq.(\ref{C7}) as the correct result for $P(\overline{Ok},{Ok})$.
\begin{figure}[tb]
\includegraphics[angle=0,width=16cm]{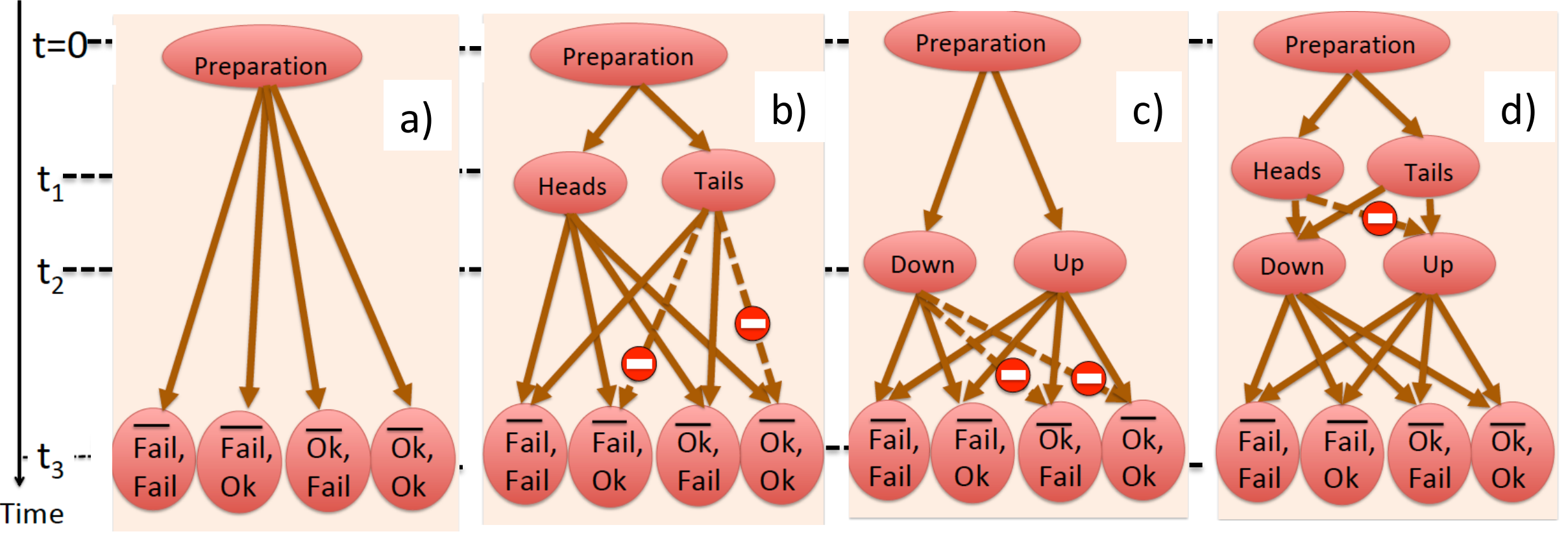}
\caption{Real paths connecting the outcomes the available 
outcomes if a) both $\x F$'s and $F$'s records are preserved;
b) only $\x F$'s record is preserved 
c) only $ F$'s record is preserved 
and both $\x F$'s and $F$'s records are destroyed
The probabilities for the paths shown
by the dashed lines vanish.}
\label{fig:FIG2}
\end{figure}
\newline 
Note that the rule D) of Sect. II
(which represents here the Uncertainty Principle \cite{FeynL}) 
forbids making
any assumptions about whether the coin was "heads" or "tails" at $t=t_{1}$,
or if the spin was "up" or "down at $t=t_{2}$. 
Note also that one still can inspect, for example, the state of the $\x F$'s record after
$\x W$ has completed his measurement. However, just as in the example at the end of 
Appendix B, finding it in a state $|D(heads)\ra$ cannot be taken as a proof that 
the coin was heads up at $t=t_1$.
\newline
\subsection{Only $\overline F$'s record is preserved}
Next assume that $\x W$  measures in the basis (\ref{pros1}) and  preserves $\x F$'s record,  
while $W$  erases the record of $F$. 
Now, in Fig.1,  only the paths paths passing through the  states $|heads\ra$ and $|tails\ra$
can be distinguished by consulting the existing records at the end of experiment. 
The paths arriving at a final state via $|tails{\rangle}|down{\rangle}
$ and $|tails{\rangle}|up{\rangle}$ at $t=t_{2}$ cannot be told apart, and,
following the rules of Sec. \ref{res}, their amplitudes should be added.
There are eight real paths connecting the observed outcomes (see Fig. 2b),
and a set of eight probabilities, 
\begin{eqnarray}\label{onlyFbar}
P^{\prime}(\overline{Fail},Fail,Heads) & =|A_{1}|^{2}=1/12,\quad
P^{\prime}(\overline{Fail},Fail,Tails)=|A_{2}+A_{3}|^{2}=1/3\quad
 \\
P^{\prime}(\overline{Fail},Ok,Heads) & =|A_{4}|^{2}=1/12,\quad
P^{\prime}(\overline{Fail},Ok,Tails)=|A_{5}+A_{6}|^{2}=0,\quad\quad
\quad  \notag \\
P^{\prime}(\overline{Ok},Fail,Heads) & =|A_{7}|^{2}=1/12,\quad
P^{\prime}(\overline{Ok},Fail,Tails)=|A_{8}+A_{9}|^{2}=1/3,\quad \quad
\notag \\
P^{\prime}(\overline{Ok},Ok,Heads) & =|A_{10}|^{2}=1/12,\quad
P^{\prime}(\overline{Ok},Ok,Tails)=|A_{11}+A_{12}|^{2}=0.\quad\quad
\quad  \notag
\end{eqnarray}
Now the joint probability for $W$ to obtain an "${Ok}$" and for $\overline{F}
$ to see "$Tails$" is zero, $P^{\prime}({Ok},Tails)=P^{\prime}(%
\overline{Fail},{Ok},Tails)+P^{\prime }(\overline{Ok},{Ok},Tails)=0$
and we recover the result (\ref{C2}) in Sect. IV.
Finally, having learnt  that $W$'s result is "$Ok$" 
we also know that $F$ had seen "$Heads$", i.e., obtain the condition (ii) of  Sect. IV,
\begin{equation}
{Ok}\rightarrow Heads.   \label{COND3}
\end{equation}
\subsection{Only $F$'s record is preserved}
If $W$ chooses the basis  (\ref{pros2}),
one is unable to distinguish between
the virtual paths in Fig. 1, which arrive at the same final state via $|heads{%
\rangle}|down{\rangle}$ and $|tails{\rangle}|down{\rangle}$ at $t=t_{1}$.
As before, there are eight real paths connecting the observed outcomes  (see Fig. 2c), and
the eight new probabilities are given by 
\begin{eqnarray}\label{onlyF}
P^{\prime\prime}(\overline{Fail},Fail,Up) & =|A_{3}|^{2}=1/12,\quad
P^{\prime\prime}(\overline{Fail},Fail,Down)=|A_{1}+A_{2}|^{2}=1/3,\quad
\label{r6a-F} \\
P^{\prime\prime}(\overline{Fail},Ok,Up) & =|A_{6}|^{2}=1/12,\quad
P^{\prime\prime\prime}(\overline{Fail},Ok,Down)=|A_{4}+A_{5}|^{2}=1/3,\quad%
\quad  \notag \\
P^{\prime\prime}(\overline{Ok},Fail,Up) & =|A_{9}|^{2}=1/12,\quad
P^{\prime\prime\prime}(\overline{Ok},Fail,Down)=|A_{7}+A_{8}|^{2}=0,\quad
\quad\quad  \notag \\
P^{\prime\prime}(\overline{Ok},Ok,Up) & =|A_{12}|^{2}=1/12,\quad
P^{\prime\prime}(\overline{Ok},Ok,Down)=|A_{10}+A_{11}|^{2}=0.\quad
\quad\quad  \notag \\
&  \notag
\end{eqnarray}
From this we learn that the odds on $F$ and $\overline{W}$ seeing "$Down$"
and "$\overline{Ok}$", respectively, are zero, $P^{\prime\prime}(%
\overline{Ok},Down)=P^{\prime\prime}(\overline{Ok},{Fail}%
,Down)+P^{\prime\prime}(\overline{Ok},{Ok},Down)=0$, recover the result (\ref{C4}). Therefore,
knowing that $\overline{W}$ obtains "$\overline{Ok}$", we also know that $F$%
's result was "$Up$". This is the condition (i) of Sect. IV,
\begin{equation}
\overline{Ok}\rightarrow Up.   \label{COND2}
\end{equation}
\subsection{Both $\bar{F}$'s and $F$'s records are preserved}
If  $W$ and $\x{W}$
measure in the bases (\ref{pros1}) and (\ref{pros2}), respectively, 
and leave $\bar{F}$'s  and $F$'s records untouched,
all paths can be distinguished when the experiment is finished. 
Each path can be endowed with a probability 
$|A_i|^2$, $i=1,...,12$ and we write down the $12$ possible outcomes in a compact form as 
\begin{eqnarray}\label{all}
P(N,M,K,I)  =1/12,\quad M,K\neq Up,Heads \n  
P(N,Up,Heads,I)  = 0,\q\q\q\q\q\q\q\q 
\end{eqnarray}
where 
\begin{equation}
I=Heads,Tails,\quad K=Up,Down,\quad M=\overline{Ok},\overline{Fail},\quad
N=Ok,Fail.   \label{r5a}
\end{equation}
The network of the {real} pathways connecting possible outcomes is shown in Fig.
2d.  The joint probability of $\overline{F}$ and $F$ seeing "$Heads$"
and "$Up$" vanishes since in Fig.2d no pathway connects the
outcomes $Heads$ and $Up$.
We also have 
 \begin{eqnarray} \label{w8aa}
P(Down,Heads)= \sum_{M,N} P(N,M,Down,Heads)=1/3, \n
P(Up,Tails)= \sum_{M,N}  P(N,M,Up,Tails)=1/3,\n
P(Down,Tails)= \sum_{M,N}  P(N,M,Down,Tails)=1/3,
 \end{eqnarray}
and recover Eqs.(\ref{C2}) and (\ref{C3}), as well as  the condition (iii) of Sect. IV,
\begin{equation}
{Up}\rightarrow{Tails}.   \label{COND1}
\end{equation}
This is the only scenario in which the measurement records of all four agents are 
retained, and the joint probabilities for all outcomes exist.
\section{Discussion and conclusions \label{conclu}}
Consider an experiment consisting of measuring, one after another, $L$ quantities, $\mathcal Q^\ell$, 
taking certain discrete values $Q^\ell_{m_\ell}$. If the system under consideration is stochastic, i.e., none of the 
values $Q^\ell_{m_\ell}$ can be uniquely deduced from other outcomes, the experiment 
relies on recording the measured values, an keeping the records until the last value, $Q^L_{m_L}$, 
is measured. 
Individual outcomes can be written down on a piece of paper, memorised, or encoded into any material object.
In the end, the records are compared, and a result $\{ Q^L_{m_L}, ...Q^\ell_{m_\ell},...Q^1_{m_1}\}$
is established. Repeating the procedure many times, one evaluates the probabilities for all possible results,
$P( Q^L_{m_L }...\gets Q^\ell_{m_\ell}...\gets Q^1_{m_1})$.
If a particular record, say, that of a value $Q^\ell_{m_\ell}$ is destroyed or erased in each run, the lost value cannot be
recovered, and the experiment proceeds as it has not been measured at all. 
\newline
Quantum systems are by their very nature stochastic, and all that was just said applies also  to successive quantum measurements. 
There is, however, one important difference. If classically one fails to record, or to keep the record of a value $Q^\ell_{m_\ell}$,
 the measured probability distribution, $P( Q^L_{m_L ...}\gets Q^{\ell+1}_{m_{\ell+1}},Q^{\ell-1}_{m_{\ell-1}}...\gets Q^1_{m_1})$,  is a marginal of the distribution obtained in a more detailed experiment,  
$$P_{class}( Q^L_{m_L ...}\gets Q^{\ell+1}_{m_{\ell+1}},Q^{\ell-1}_{m_{\ell-1}}...\gets Q^1_{m_1})= \sum_{m_\ell}P_{class}( Q^L_{m_L }...\gets Q^\ell_{m_\ell}...\gets Q^1_{m_1}).$$
In general, this rule does not hold in the quantum case.
This \e{only mystery of quantum mechanics} was illustrated by Feynman \cite{FeynL} on the simplest case of the Young's double slit experiment, 
where the failure to produce a record of a slit taken by an electron, or its subsequent destruction \cite{WFS-EPL}, leads to the appearance
of an interference pattern, otherwise absent. The same principle applies to more complex situations, such as the one 
considered in this paper. Quantum mechanics does not hesitate to predict probabilities in all practical situations. 
However,  the resulting statistical ensembles depend on the retained measurement records, and are not guaranteed to be compatible.
\newline
{\color{black}
Indeed, let us look at the  three conflicting statements of Section V.
The last of them iii), is obtained under the assumption that
both $\x F$ and $F$ make their measurements, and their records are preserved until the end of the experiment.
The other two conditions,  ii) and i), correspond to the cases where the record of one 
of the Friends, $\x F$ or $F$, is either erased by $\x W$'s or $W$'s.
Each claim
  is correct under the
proper circumstances, and one is never in doubt as to which one is to be used. 
The Feynman rules specify, depending on the measurements that are effectively recorded, whether one should sum over virtual paths (amplitudes) or over real paths (probabilities).
\newline
It is also easy to see what caused the apparent contradiction in Sect.V.
The unitary evolved state (\ref{C1}) contains no information about $\x W$'s and $W$'s arrangements, 
and must cater for all eventualities. 
Having stipulated that all record of $F$'s and $\x F$'s measurements are to be erased by the actions of $W$ and $\x W$, 
one should go straight to Eq.(\ref{C7}). 
According to the Feynman's rules, there is no contradiction.
Either $\W$ and $W$ measure the entire laboratories, thus erasing $\x F$'s and $F$'s records, 
or preserve the records by measuring only a part of a lab.  One cannot have both situations at the same 
time 
  \br{without violating the rules. This would then imply that measurement outcomes would still be defined although all the corresponding material records in the Universe (including in F’s brain) have been erased.}}
\newline
In summary, quantum mechanics can  be wrong in its predictions, 
but it is not inconsistent, at least as far as the Wigner's friend scenarios discussed here are concerned.
It is, of course, still legitimate to question
Feynman's  principle by anticipating situations in which 
they would fail, or where different interpretations,  such as used 
in \cite{wigner}-\cite{wiseman} would make verifiable 
alternative  predictions, one would need a stronger case to make further progress in the direction. 
\appendix

\section{Application of Feynman's rules \label{AFR}}

In order to apply the principles (A-F) given in Sec. \ref{res} 
, the following steps are required:\newline
a) Decide which quantities are to be measured, and at what times. With each
chosen quantity associate a hermitian operator, whose eigenvalues will be
a possible measured outcomes. The first measurement must prepare the
composite in a known initial state.\newline
b) Define the bases needed for all measurements in the Hilbert space of a
composite system which includes all subsystems interacting with each other in
the course of the experiment.  Connect all states, to obtain all
virtual paths. \newline
c) The probability amplitude of a path is given by a product of the matrix
elements of the evolution operators between the times of two consecutive
observations. \newline
d) Decide which paths will interfere, depending on the degeneracies of the
eigenvalues obtained in the intermediate measurements. Add up the amplitudes
for the paths leading to the same final state, and take the absolute square, 
\newline
e) Sum the probabilities obtained in (d) over the degeneracies of the last
eigenvalue. 
\newline
It is important to bear in mind that different sets of measurements can
produce {different (incompatible)} statistical ensembles, even if
they are made on the same system, and even in the case when the measured
operators commute, but have a different sets of degenerate eigenvalues (see,
e.g., \cite{DSann}). 
\newline
Finally, the probabilities one calculates ought to be the likelihoods of the
outcomes experienced by all Observers taking part in the experiment. Such
experiences may be gained by registering the conditions of the devices
(probes) accessible to an Observer at the end of experiment, or by
consulting Observer's s memory, which carries a record of a previously registered outcome 
\cite{WFS-EPL}, \cite{DSent}. 
\section{Appendix B: a double-slit example}
In \cite{FeynL} Feynman illustrated his basic principles of quantum motion on a double-slit experiment, 
which we will revisit next, albeit is a slightly different form.
where a photon is scattered by electron into one of  two distinguishable states, depending 
on the slit chosen by the electron. The presence of such a photon allows, in principle, to determine which slit
was used, and the interference pattern on the screen is destroyed even if the photon is {\it never observed}. 
\newline
The simplest double-slit problem can be conceived  by considering a two-level system, $S$, and two probes, 
$D$ and $\D$.
The probes will have no own dynamics, and move only as a result of coupling with other parts of the joint system. 
 At $t=t_0$, the three are prepared a state $|\Psi_0\ra=|\D(0)\ra \ot |D(0)\ra\ot |s_0\ra$, 
and at $t_1>t_0$ and at $t_2>t_1$ the two probes coupled to the system according to ($|s\ra$ is an arbitrary system's state, 
and $|up\ra$, $down\ra$ and $|fail\ra$, $ok\ra$ are two orthogonal bases in its Hilbert space)
 \begin{eqnarray} \label{w1}
\u^{D+S}(t_1)|D(0)\ra\ot |s\ra=\la up|s\ra |D(up)\ra \ot |up\ra+\la down|s\ra |D(down)\ra \ot |down\ra, \n
\u^{\D+S}(t_2)|\D(0)\ra\ot |s\ra=\la fail |s\ra |\D(fail)\ra \ot |fail\ra+\la ok|s\ra |\D(ok)\ra \ot |ok\ra,
 \end{eqnarray}
 where
  \begin{eqnarray} \label{w1a}
|fail\ra=\alpha |up\ra + \beta |down\ra, \q |ok\ra=\gamma |up\ra + \delta |down\ra.
 \end{eqnarray}
Unitary evolution of  $|\Psi_0\ra$ to a time just after $t_2$  yields
  \begin{eqnarray} \label{w1b}
|\Psi(t)\ra\equiv u^{\D+S}(t_2)\u^S(t_2,t_1)\u^{D+S}(t_1)\u^S(t_1,t_0)|\D(0)\ra \ot |D(0)\ra\ot |s_0\ra\q\n
=\sum_{i,j}A^S(j\gets i \gets s_0)|\D(j)\ra\ot |D(i)\ra\ot |j\ra,\q\q\q\q\q\q\q\q \n
A^S(j\gets i \gets s_0) \equiv \la j|\u^S(t_2,t-1)|i\ra \la i|\u^S(t_1,t_0)|s_0\ra,\q i=up,down,\q j=fail,ok,
 \end{eqnarray}
where $A^S(j\gets i \gets s_0)$ is the amplitude of a system's virtual path $\{j\gets i \gets s_0\}$,
and
$\u^S$ is the system's evolution operator. The  joint probability of $F$ and $W$ seeing the outcomes $i$ and $j$,
 \begin{eqnarray} \label{w2}
P(i,j) \equiv \la\Psi(t)|\D(j)\ra\la \D(j)|\ot |D(i)\ra\la D(i)|\Psi(t)\ra =|A^S(j\gets i \gets s_0)|^2,
 \end{eqnarray}
and, in particular, the odds on $W$ seeing an outcome $Ok$ is
   \begin{eqnarray} \label{w3}
P(ok)\equiv P(ok,up)+P(ok,down) = |A^S(ok\gets up \gets s_0)|^2+|A^S(ok\gets down \gets s_0)|^2\q
 \end{eqnarray}
 \newline
In a different scenario, $F$ may decide to to measure, so his probe, not coupled to the system, continues in its initial  state $|D(0)\ra$.
Now the unitary evolution yields
 becomes
   \begin{eqnarray} \label{w4}
|\Psi'(t)\ra\equiv u^{D+S}(t_2)\u^S(t_2,t_0)|\D(0)\ra \ot |D(0)\ra\ot |s_0\ra=\q\q\q\q\n
[A^S(fail\gets up \gets s_0)+A^S(fail\gets down \gets s_0)]|\D(fail)\ra\ot  |fail\ra\ot |D(0)\ra\n
+[A^S(ok\gets up \gets s_0)+A^S(ok\gets down \gets s_0)]|\D(ok)\ra\ot |ok\ra\ot |D(0)\ra,\q 
 \end{eqnarray}
We, therefore have  
    \begin{eqnarray} \label{w5}
P'(ok)\equiv \la\Psi(t)|\D(ok)\ra\la \D(ok)|\Psi(t)\ra= |A^S(ok\gets up \gets s_0)+A^S(ok\gets down \gets s_0)|^2,\q
 \end{eqnarray}
and need not examine the first probe, decoupled from the system of interest. 
\newline 
 Comparing Eqs.(\ref{w2}) and (\ref{w4}) one makes two useful observations.
Firstly, if only von Neumann couplings are applied, the probability of interest can be expressed 
via probability amplitudes evaluated for the studied system in the absence of the probes. 
This was first noted by von Neumann \cite{vN}. 
\newline
Secondly, the probabilities can be constructed according to the Feynman's rules applied to the studied 
system only, whose virtual paths interfere if $D$ is not engaged, but become exclusive \e{real}
alternatives if $D$ is coupled, and the paths can be distinguished.
\newline
One thing that can be added to this simple narrative is this \cite{WFS-EPL}. The Observer may set up the second 
probe to measure a composite $\{system + D\}$
The simplest double-slit problem can be conceived  by considering a two-level system, $S$, and two probes, 
$D$ and $\D$ in a basis (the evolution will occur in a two dimensional 
sub-space of the four-dimensional Hilbert space, and we will not bother to specify all four basis states)
  \begin{eqnarray} \label{w5a}
|Fail\ra=\alpha |D(up)\ra \ot |up\ra + \beta|D(down)\ra \ot |down\ra, \n
|Ok\ra=\gamma |D(up)\ra \ot |up\ra + \delta|D(down)\ra \ot |down\ra.\
 \end{eqnarray}
The new coupling of $W$'s probe, therefore, is  ($|\Phi\ra$ is any state of the composite in the sub-space spanned 
by $|Fail\ra$ and $|Ok\ra$)
 \begin{eqnarray} \label{w6}
\u^{\Pr+Pr+S}(t_2)|\D(0)\ra\ot |\Phi\ra=
\la Fail |\Phi\ra |\D(Fail)\ra \ot |Fail\ra+\la Ok|\Phi\ra |\D(Ok)\ra \ot |Ok\ra,\q
 \end{eqnarray}
 and for the unitarily evolved state, just after $t_2$, we find
    \begin{eqnarray} \label{w7}
|\Psi''(t)\ra\equiv \u^{\Pr+Pr+S}(t_2)\u^S(t_2,t_1)\u^{Pr+S}(t_1)\u^S(t_1,t_0)|\D(0)\ra \ot |D(0)\ra\ot |s_0\ra=\q\q\n
[A^S(fail\gets up \gets s_0)+A^S(fail\gets down \gets s_0)]|\D(Fail)\ra\ot  |Fail\ra\q\q\q\q\n
+[A^S(ok\gets up \gets s_0)+A^S(ok\gets down \gets s_0)]|\D(Ok)\ra\ot |Ok\ra, \q\q\q\q\q 
 \end{eqnarray}
 so that
  \begin{eqnarray} \label{w9}
P''(Ok)\equiv  = \la\Psi''(t)|\D(Ok)\ra\la \D(Ok)||\Psi''(t)\ra=\q\q\q\n
|A^S(ok\gets up \gets s_0)+A^S(ok\gets down \gets s_0)|^2=P'(ok).
 \end{eqnarray}
 As before, the probabilities of interest can be obtained by manipulating 
the system's path amplitudes,  $A^S$, and the result still agrees with the Feynman's rules of Sect.2A. 
The application of the second probe [cf.  (\ref{w6})] has {\it erased}, or if one prefers
 {\it destroyed}, the record carried by the first probe, the system's paths passing through the states 
 $|up\ra$ or $|down\ra$ at $t=t_1$ cannot be distinguished, and their amplitudes must be added. 
 \newline
 Note that one can still check whether the first probe is in a state $|D(up)\ra$ at some $t_3>t_2$, 
 e.g., just after $t_2$, so that $\u^S(t_3,t_2)=1$.
In this case, using the Feynman's rules (in the Hilbert space of $\{S+D\}$) and evaluating the matrix elements, one  easily finds
 \begin{eqnarray} \label{w8a}
P'''(up, Ok) =\sum_{i=up,down}|A^{S+D}(up\ot D(i)\gets Ok \gets D(0)\ot s_0)|^2\n
=|A^{S+D}(up\ot D(up)\gets Ok \gets D(0)\ot s_0)|^2
=|A^{S}(up \gets ok\gets up \gets s_0)|^2 \n
=|\gamma|^2 |A^{S}(ok\gets up \gets s_0)+A^{S}(ok\gets down \gets s_0)|^2 =
 \end{eqnarray} 
and 
 \begin{eqnarray} \label{w8aa}
P'''(up, Fail) 
=|\alpha|^2 |A^{S}(ok\gets up \gets s_0)+A^{S}(ok\gets down \gets s_0)|^2.
 \end{eqnarray} 
 Finally, since $\la j'|j\ra=\delta_{j'j}$,  $j,j'=Fail, Ok$, we have $\gamma=\beta^*$, so that
 \begin{eqnarray} \label{w8}
P'''(up) = P'''(up,Ok)+P'''(up,Fail) =
|A^S(ok\gets up \gets s_0)+A^S(ok\gets down \gets s_0)|^2,\q\q\q
 \end{eqnarray} 
 and, as expected, the information about the system's state at $t_1$ is lost to interference. 
 \newline

\begin{center}
\textbf{Acknowledgements}
\end{center}
Financial support of
MCIU, through the grant
PGC2018-101355-B-100(MCIU/AEI/FEDER,UE)  and the Basque Government Grant No IT986-16,
is acknowledged by DS.

\end{document}